\documentstyle[12pt]{article}   

\large
\newcommand{\beq}{\begin{equation}}
\newcommand{\eeq}{\end{equation}}

\makeatletter
\@addtoreset{equation}{section}
\makeatother

\newtheorem{Theorem}{Theorem}[section]
\newtheorem{Definition}{Definition}[section]
\newtheorem{Lemma}{Lemma}[section]

\def\Cyl{{\rm Cyl}}

\def\be{\begin{equation}}
\def\ee{\end{equation}}
\def\ba{\begin{eqnarray}}
\def\ea{\end{eqnarray}}

\def\ag{{{\cal A}/{\cal G}}}

\def\agb{{\overline {{\cal A}/{\cal G}}}}

\def\Comp{{\mathchoice
{\setbox0=\hbox{$\displaystyle\rm C$}\hbox{\hbox to0pt
{\kern0.4\wd0\vrule height0.9\ht0\hss}\box0}}
{\setbox0=\hbox{$\textstyle\rm C$}\hbox{\hbox to0pt
{\kern0.4\wd0\vrule height0.9\ht0\hss}\box0}}
{\setbox0=\hbox{$\scriptstyle\rm C$}\hbox{\hbox to0pt
{\kern0.4\wd0\vrule height0.9\ht0\hss}\box0}}
{\setbox0=\hbox{$\scriptscriptstyle\rm C$}\hbox{\hbox to0pt
{\kern0.4\wd0\vrule height0.9\ht0\hss}\box0}}}}
\def\Co{{\mathchoice
{\setbox0=\hbox{$\displaystyle\rm C$}\hbox{\hbox to0pt
{\kern0.4\wd0\vrule height0.9\ht0\hss}\box0}}
{\setbox0=\hbox{$\textstyle\rm C$}\hbox{\hbox to0pt
{\kern0.4\wd0\vrule height0.9\ht0\hss}\box0}}
{\setbox0=\hbox{$\scriptstyle\rm C$}\hbox{\hbox to0pt
{\kern0.4\wd0\vrule height0.9\ht0\hss}\box0}}
{\setbox0=\hbox{$\scriptscriptstyle\rm C$}\hbox{\hbox to0pt
{\kern0.4\wd0\vrule height0.9\ht0\hss}\box0}}}}
\def\Rl{{\mathchoice
{\setbox0=\hbox{$\displaystyle\rm R$}\hbox{\hbox to0pt
{\kern0.4\wd0\vrule height0.9\ht0\hss}\box0}}
v{\setbox0=\hbox{$\textstyle\rm R$}\hbox{\hbox to0pt
{\kern0.4\wd0\vrule height0.9\ht0\hss}\box0}}
{\setbox0=\hbox{$\scriptstyle\rm R$}\hbox{\hbox to0pt
{\kern0.4\wd0\vrule height0.9\ht0\hss}\box0}}
{\setbox0=\hbox{$\scriptscriptstyle\rm R$}\hbox{\hbox to0pt
{\kern0.4\wd0\vrule height0.9\ht0\hss}\box0}}}}

\title{A length operator for canonical quantum gravity}
\author{T. Thiemann\thanks{thiemann@math.harvard.edu} \\
       Physics Department, Harvard University, \\
       Cambridge, MA 02138, USA}
\date{{\small Preprint HUTMP-96/B-354}}

\begin{document}

\maketitle

\begin{abstract}
We construct an operator that measures the length of a curve in 
four-dimensional Lorentzian vacuum quantum gravity.

We work in a representation in which a $SU(2)$ connection is diagonal
and it is therefore surprising that the operator obtained after 
regularization is densely defined, does not suffer from factor ordering 
singularities and does not require any renormalization.

We show that the length operator admits self-adjoint extensions and 
compute part of its spectrum which like its companions, the volume and 
area operators already constructed in the literature, is purely discrete
and roughly is quantized in units of the Planck length.

The length operator contains full 
and direct information about all the components of the metric tensor 
which faciliates the construction of a new type of weave states which 
approximate a given classical 3-geometry.

\end{abstract}

\section{Introduction}

If one was working in a representation of canonical quantum gravity for
which the intrinsic metric $q_{ab}$ of an initial data 
hypersurface was the configuration variable then the operator corresponding
to the length of a curve would be fairly easy to construct because it
would act by multiplication.\\
However, in this so-called geometrodynamic representation \cite{1} one of 
the most
important operators, the Wheeler-DeWitt constraint operator, adopts an
algebraic form which is so difficult that almost half a century 
after its discovery it is still unknown how to rigorously define it
and much less how to solve it (compare \cite{2} for a detailed analysis). \\
Fortunately, there is an alternative, called the connection representation,
in which the Wheeler-DeWitt operator adopts a polynomial form after 
multiplying it by $\sqrt{\det((q_{ab}))}$. In the Lorentzian signature
the underlying connection is complex valued \cite{3} which renders the task 
of incorporating the correct reality conditions on the classical phase
space variables at the quantum level into a difficult one. However,
these complications can be overcome in two different ways : the first
possibility is to start with Euclidean gravity, in which the Wheeler-DeWitt
constraint is also a polynomial \cite{4}, and to Wick rotate the Euclidean
theory into the Lorentzian regime \cite{5}. The second possibility arises
from the recent discovery \cite{6,7} that by employing a certain novel 
technique it
is indeed possible to obtain a finite, densely defined (and symmetric)
operator
corresponding to the original Wheeler-DeWitt constraint directly in the 
Lorentzian
regime and without multiplying by a power of $\sqrt{\det(q)}$. This
technique is independent of the one described in \cite{8} which was 
restricted to the classical case and unavoidably is doomed to produce
a highly singular Hamiltonian constraint operator 
(compare \cite{21} where this singularity is discussed in the restricted 
context of lattice quantum gravity). In contrast, the one proposed 
in \cite{7,8} is perfectly finite everywhere on the Hilbert space.
Also, it is independent of the technique used in \cite{9} which suffers from
other problems which have to do with taking the square root of a non-positive
and not self-adjoint operator besides being a quantization of the Euclidean
constraint operator only.\\
In the real connection representation which we will use in the sequel the 
metric depends on the momentum operator only, and it depends on it in 
a non-polynomial fashion. It is therefore much harder to define a 
well-defined operator corresponding to it. In fact, that is why only volume
and area operators could be constructed so far \cite{10,11}. It turns
out that the same novel technique introduced in \cite{6,7} can be used to
derive a completely well-defined operator corresponding to the length of a
curve.\\
The article is organized as follows :

In section 2 we fix the notation and recall the necessary information about
the Hilbert space and techniques that come with the space of generalized 
connections modulo gauge transformations $\agb$.

In section 3 we derive the length operator and show that it is an unbounded
symmetric operator on the full Hilbert space with dense domain and that 
it has self-adjoint extensions.

In section 4 we derive several properties of its spectrum and compute it for
some simple situations. 

In section 5 we comment on how one can construct ``weave states" 
which approximate a given classical geometry. We need to consider weave states
which are more general than most of the ones previously considered in the 
literature in the sense that they necessarily involve intersections and 
overlappings of the loops involved and should therefore be rather 
called ``lattice states".

\section{The real connection representation}

Let $q_{ab}$ be the intrinsic metric of an initial data hypersurface $\Sigma$
and let $K_{ab}$ be its extrinsic curvature. Introduce a triad field 
$e_a^i$ which transforms like a 1-form under diffeomorphisms of $\Sigma$ and
according to the adjoint representation of $SU(2)$ so that 
$q_{ab}=\delta_{ij}e_a^i e_b^j$. Let $e^a_i$ 
be its inverse and define $K_a^i:=\mbox{sgn}(\det((e_b^j)))K_{ab}e^b_i$
and $E^a_i:=\det((e_b^j))e^a_i$. Then one can show that $(K_a^i/\kappa,
E^a_i)$ is a canonical pair for Lorentzian four-dimensional canonical
gravity, where $\kappa$ is Newton's constant.\\
Now consider the spin-connection $\Gamma_a^i$, that is, the unique 
connection which annihilates the triad $e_a^i$. One can define 
a $SU(2)$ connection $A_a^i:=\Gamma_a^i+K_a^i$ which has the correct
dimension of an inverse length. Then one can show that 
the so-called real Ashtekar variables given by $A_a^i/\kappa, E^a_i$
define a canonical pair. This observation is due to Ashtekar \cite{3,4}.
In retrospect, since this real connection formulation does not simplify
the Wheeler-DeWitt constraint too much while the complex formulation 
suffers from the other problems mentioned in the introduction, the virtue of 
using a connection dynamics 
formulation rather than a geometrodynamical one is the following :
one can use techniques normally employed in Yang-Mills theory.
In particular, one can use Wilson loop variables which serve as good
coordinates on $\ag$, the space of smooth connections modulo gauge 
transformations. The use of loops to probe connections is radical :
those Wilson loop variables can become well-defined operators only if 
the excitations of geometry are string-like rather than bubble-like.
On the other hand, given that assumption, it is possible to explicitly
characterize the quantum configuration space $\agb$ of {\em generalized}
connections modulo gauge transformations. This is the precise analogue,
in the connection representation, of ``Wheeler's superspace" in the metric
formulation which, to the best of our knowledge, was never specified 
precisely.\\
One can show that the elements of $\agb$ are in one-to-one correspondence
with the set of all homomorphisms from the group of holonomically equivalent
loops in $\Sigma$ into $SU(2)/\mbox{Ad}$ \cite{12}. Moreover, 
there is a $\sigma$ -- additive, faithful and diffeomorphism invariant 
probability measure $\mu_0$ on $\agb$ which equips us with a Hilbert 
space structure ${\cal H}:=L_2(\agb,d\mu_0)$. This measure is defined 
as follows :\\
In the sequel we will denote by $\gamma$ a closed graph, that is, a collection 
of analytic edges $e$ which intersect in vertices $v$ such that each vertex 
is at least two-valent. A function $f$ on $\agb$ is said to be cylindrical
with respect to $\gamma$ iff it is of the form 
$f(A)=(f_\gamma\circ p_\gamma)(A)=f_\gamma(h_{e_1}(A),..,h_{e_n}(A))$ where 
$e_1,..,e_n$ are the 
edges of $\gamma$, $h_e$ is the holonomy along $e_i$ and $f_\gamma$ is a
complex-valued function on $SU(2)^n$ which is gauge invariant. 
So each cylindrical function is determined through a graph $\gamma$ and 
such an $f_\gamma$ and one says that $f_\gamma$ and $f_\gamma'$ are 
equivalent whenever their pullbacks agree, that is $p_\gamma^\ast 
f_\gamma=p_{\gamma'}^\ast f_{\gamma'}$. Let us 
denote by $\mbox{Cyl}_\gamma(\agb)$ the collection of functions 
cylindrical with respect to $\gamma$ modulo cylindrical equivalence and 
denote by
$\mbox{Cyl}(\agb):=\cup_\gamma\mbox{Cyl}_\gamma(\agb)$ the set of all
cylindrical functions.\\
The measure $(\mu_0)$ can now be seen as the $\sigma$ -- additive extension
\cite{13} of the following self-consistent family of measures 
$(\mu_{0,\gamma})_\gamma$ \cite{14} : let $f\in\Cyl_\gamma(\agb)$ then
\be \label{1}
\int_\agb d\mu(A) f(A):=\int_\agb 
d\mu_{0,\gamma}(A) f(A):=\int_{SU(2)^n}d\mu_H(g_1)..d\mu_H(g_n) 
f_\gamma(g_1,..,g_n) \ee
where $\mu_H$ is the Haar measure on $SU(2)$ and $g_I=h_{e_I}(A)$. In other
words, everything is reduced to finite dimensional integrals over 
$SU(2)$. One can show that $\mbox{Cyl}(\agb)$ is dense in $\cal H$.\\
Thus integral calculus is introduced on $\agb$. One can even develop 
differential calculus on $\agb$ \cite{15} : a differentiable cylindrical
function is simply a differentiable function on $SU(2)^n$. Functional
derivatives can be evaluated on differentiable elements of $\Cyl(\agb)$ 
because 
on a finite graph a distributional connection can be replaced by a smooth 
one \cite{12,14}. One sees that differential and integral calculus is
inherited from the one on $SU(2)^n$. Finally, let us define the spaces 
$\mbox{Cyl}_\gamma^n(\agb),\;\mbox{Cyl}^n(\agb)=\cup_\gamma 
\mbox{Cyl}_\gamma^n(\agb)$ of differentiable cylindrical functions of
order $n=0,1,..,\infty$. Each of $\mbox{Cyl}^n(\agb)$ is dense in $\cal H$.

\section{The length operator}

The stage is now set to derive the length operator. Before we do that
we wish to comment why it was up to now impossible to define this
operator rigorously on $\cal H$ and why it was possible to define 
volume and area operators.\\
Consider for instance the volume of a region $R\subset\Sigma$
\be \label{2}
V(R):=\int_R d^3x \sqrt{|\det(q)|}=\int_R d^3x \sqrt{|\frac{1}{3!}
\epsilon_{abc}\epsilon^{ijk}E^a_i E^b_j E^c_k|}
\ee
where $\epsilon_{abc}$ carries density weight $-1$. We see that this 
functional involves the square root of a {\em polynomial} in $E^a_i$.
This is important because, according to the canonical commutation 
relations, we are supposed to replace $E^a_i$ by $\hat{E}^a_i=-i\ell_p^2
\frac{\delta}{\delta A_a^i}$ where $\ell_p:=\sqrt{\kappa\hbar}$ is the 
Planck length. Now one tries to regularize the polynomial and defines its
square root by its spectral resolution. This has been done successfully
in the literature \cite{10,11}.\\
We know how to regularize a 
polynomial in the basic operators $A,E$ but certainly we do not know 
how to define a non-polynomial function. This is precisely the case for
\be \label{3}
q_{ab}=\epsilon_{acd}\epsilon_{bef}\epsilon^{ijk}\epsilon^{imn}
\frac{E^c_j E^d_k E^e_m E^f_n}{4\det((E^g_l))^2} \;.
\ee
Even if one could define it, it is by now well known that the operator
version of the denominator of (\ref{3}) has a huge kernel so that
it could not be defined on a dense domain. 
This is the reason why a quantization of the length of a piecewise 
differentiable curve $c\;:\; [0,1]\to\Sigma\; ;\; t\to c(t)$
given classically by 
\be \label{4}
L(c):=\int_{[0,1]} dt\sqrt{\dot{c}^a(t)\dot{c}^b(t) q_{ab}(c(t))}
\ee
has escaped its quantization in the representation $\cal H$ so far.
Note that no absolute value signs are necessary in (\ref{4}) under the 
square root since $\dot{c}^a\dot{c}^b q_{ab}=
(\dot{c}^a e_a^i)(\dot{c}^b e_b^j)\delta_{ij}\ge 0$.

\subsection{Regularization of the length operator}

The regularization of (\ref{4}) is based on two {\em key} observations :\\
Observation 1)\\
The triad is integrable with generating functional $V:=V(\Sigma)$, the total
volume of $\Sigma$.
More precisely we have $e:=\mbox{sgn}(\det((e_a^i)))$
\be \label{5}
\frac{\delta V}{\delta E^a_i}=e\frac{e_a^i}{2}=\frac{1}{\kappa}\{A_a^i,V\}
\ee
which can be verified immediately.\\
Observation 2)\\
The total volume can be quantized in a mathematically rigorous way. Its
action on sufficiently differentiable cylindrical functions is given by
(we follow \cite{11})
\ba \label{6}
\hat{V}f &=& \ell_p^3 \hat{v}f=\ell_p^3 \sum_{v\in V(\gamma)}\hat{V}_v 
f_\gamma\nonumber\\
&=&\ell_p^3\sum_{v\in V(\gamma)}\sqrt{|\frac{i}{8\cdot3!}
\sum_{e_I\cup e_J\cup e_K=v} 
\epsilon(e_I,e_J,e_K) \epsilon_{ijk} X_I^i X_J^j X_K^k|}\;f_\gamma\;.
\ea
Here $g_I=h_{e_I}(A),\;X_I=X(g_I)$ and $X^j(g):=\mbox{tr}(g[-i\sigma_j/2]
\partial_g)$ are the components of the right invariant vector
field on $SU(2)$ ($\sigma_j$ are Pauli matrices) 
where we have assumed that the edges are outgoing at a 
vertex. $V(\gamma)$ is the set of vertices of $\gamma$ and finally 
$\epsilon(e_I,e_J,e_K)=
\mbox{sgn}(\det(\dot{e}_I(0),\dot{e}_J(0),\dot{e}_K(0)))$. Note that 
$\hat{v}$ is a dimensionless operator. One can show that (\ref{6})
has dense domain $\mbox{Cyl}^3(\agb)$ and is an essentially self-adjoint 
operator on $\cal H$ with a purely discrete spectrum (compare the
forthcoming companion paper to \cite{11}).\\
These two observations motivate the following regularization strategy.\\
Choose the basis in $su(2)$ given by $\tau_j=-i\sigma_j/2$ 
and write $e_a=e_a^i\tau_i,\; A_a=
A_a^i\tau_i$ so that for smooth $A$ we have $h_e(A)=1+A(e)+o(A(e)^2)$ where
$A(e)=\int_e A$. Then, according to (\ref{5}) 
\be \label{7}
q_{ab}=-2\mbox{tr}(e_a e_b)
=-\frac{8}{\kappa^2}\mbox{tr}(\{A_a,V\}\{A_b,V\})\;.
\ee
Clearly we now are going to replace the integral in (\ref{4}) by a 
Riemann sum and take the limit. So let $t_0=0<t_1<..<t_n=1$ be a partition
of $[0,1]$ such that points $t$ of non-differentiability of $c$ are among
the values $t_i$ and let $\Delta_i:=t_i-t_{i-1}$. It is understood 
that in the limit $n\to\infty$ also $\delta:=\max((\Delta_i)_{i=1}^n)\to 0$.
Consider then the following quantity
\be \label{8}
L_n(c):=\frac{1}{\kappa}\sum_{i=1}^n 
\sqrt{2\mbox{tr}(\{h_c(t_{i-1},t_i),V\}\{h_c(t_{i-1},t_i)^{-1},V\})}\;.
\ee
Here $h_c(s,t)$ denotes the holonomy of $A$ along $c$ from the parameter
value $s\to t$. It is easy to see that for a classical (that is, smooth) 
connection we 
have $\{h_c(t_{i-1},t_i)^{\pm 1},V\}=\pm\Delta_i\dot{c}^a(\tilde{t}_i)
\{A_a(c(\tilde{t}_i),V\}+\mbox{o}(\Delta_i^2)$ where $\tilde{t}_i$
is some value of $t\in[t_{i-1},t_i]$. 
Therefore (\ref{8}) converges 
classically to (\ref{4}) in the limit $n\to\infty$. Note that the 
argument of the square root is manifestly gauge invariant. The motivation
to replace the $A's$ by holonomies of course comes from the fact that 
$\hat{V}$ has finite action on holonomies as seen from (\ref{6}).\\
The final step is to replace $V$ by $\hat{V}$ and Poisson brackets by
commutators times $1/i\hbar$. The result is 
\be \label{9}
\hat{L}_n(c):=\ell_p\sum_{i=1}^n 
\sqrt{-8\mbox{tr}([h_c(t_{i-1},t_i),\hat{v}][h_c(t_{i-1},t_i)^{-1},\hat{v}])}
\ee
To complete the definition of the length operator we now define for each 
thrice differentiable cylindrical function $f\in\mbox{Cyl}^3(\agb)$
\be \label{10}
\hat{L}(c)f:=\lim_{n\to\infty}[\hat{L}_n(c)f]\;.
\ee
As it stands, (\ref{10}) does not make sense yet because we have not shown
that the limit exists and it may matter how to choose the partition 
$(t_i)_{i=0}^n$. Also, it is far from obvious that the square root is 
well defined because we did not show that its argument is a positive 
and self-adjoint operator. These issues will be settled in the next 
subsections.

\subsection{Finiteness and choice of the partition adapted to a graph}

Let $f$ be a function cylindrical with respect to a graph $\gamma$ and
let $s$ be one of the segments of $c$ into which we have partitioned it. We 
wish to study the action of the operator 
\be \label{11}
\hat{l}_s^2:=-8\mbox{tr}([h_s,\hat{v}][h_s^{-1},\hat{v}])
\ee
on $f$. By choosing $n$ in (\ref{10}) large enough we may assume without 
loss of generality that $s$ and $\gamma$ are either disjoint, intersect
in at most one point or $s$ is contained in $\gamma$ (here we have made use
of the piecewise analyticity of $\gamma$ \cite{14}).
Further we adapt the partition to the graph in the 
following way : a) if $s$ and $\gamma$ intersect in a point then this point 
is a boundary point of $s$ and b) if $s$ is contained in $\gamma$ 
then it is 
contained in an edge of $\gamma$. This is a choice that we have to make in 
order to
achieve independence of the partition. The resulting operator  is 
different if one 
assumes that $s$ and $\gamma$ do not intersect in an endpoint of $s$.\\ 
We have in general 
$V(s\cup\gamma)=V(\gamma)\cup V(s)\cup V(\gamma\cap s)=V(\gamma)\cup
V(\gamma\cap s)$. The volume operator applied to a graph is a sum
of the $\hat{V}_v$ for each vertex of the graph and $\hat{V}_v$ only 
depends on those edges of the graph which are incident at $v$ as follows 
from (\ref{6}). Therefore
$h_s\hat{V}_v=\hat{V}_v h_s$ if $s$ is not incident at $v$. It follows that
\ba \label{12}
&& [h_s^{-1},\hat{v}]f=\sum_{v\in V(\gamma)} h_s^{-1}\hat{V}_v f
-\sum_{v\in V(\gamma\cup s)}\hat{V}_v h_s^{-1}f \nonumber\\
& =& \sum_{v\in V(\gamma)\cap\partial s} [h_s^{-1},\hat{V}_v] f
-\sum_{v\in V(\gamma\cap s)-V(\gamma)}\hat{V}_v h_s^{-1}f
\ea
and both sums in the last line have at most one non-vanishing term 
corresponding to an endpoint of $s$ intersecting $\gamma$.\\ 
Case 1) $\gamma\cap s=\emptyset$\\ 
Then $\hat{l}_s^2 f=0$ as is immediate from (\ref{12}).\\
Case 2) $p\in\gamma\cap s$

Subcase a) $p$ is neither a vertex of $\gamma$ nor a kink of $c$ and so 
only the second term in (\ref{12}) survives.\\
This implies that $p$ is a trivalent vertex with only two independent 
tangent directions of the edges of $\gamma$ and the segment $s$ incident at 
$p$. The properties of the volume operator now imply that the contribution 
$\hat{V}_p$ of $\hat{v}$ vanishes and therefore
$\hat{l}_p^2f=0$.

Subcase b) $p$ is not a vertex of $\gamma$ but a kink of $c$.\\
Since $s$ is  only one of the segments of $c$ incident at $p$ this
case is not different from a).

Subcase c) $p$ is a vertex of $\gamma$ but not a kink of $c$.\\
Now the result can be non-vanishing. We have for a vertex $p=v$
of $\gamma$ that
$\hat{l}_s^2f=-8\mbox{tr}([h_s,\hat{V}_v][h_s^{-1},\hat{V}_v])f$, that is,
only the first term in (\ref{12}) survives. That the volume operator in the 
second commutator 
also reduces to $\hat{V}_v$ follows by a similar argument.

Subcase d) $p$ is both a vertex of $\gamma$ and a kink of $s$.\\
Again, this case does not differ from c) for the same reason as b) was
equal to a).\\
So we conclude that for large enough $n$ the operator (\ref{9}) reduces to
\be \label{13}
\hat{L}_n(c):=\ell_p\sum_{v\in V(\gamma)}\sum_{v\in s_i}
\sqrt{-8\mbox{tr}([h_{s_i},\hat{V}_v][h_{s_i}^{-1},\hat{V}_v])}
\ee
where we have denoted the segment $s_i:=c([t_{i-1},t_i])$.
This demonstrates that $\hat{L}_n(c)$ is a finite operator for each
$n$  because there are at most $|V(\gamma)|$ terms in the sum (\ref{13}).
The next question is whether the limit $n\to\infty$ exists.

\subsection{Existence of the limit $n\to\infty$}

We now show that (\ref{13}) is actually independent of the
$s_i$ so that the limit $n\to\infty$ is already taken and the limit therefore
exists trivially.\\
In fact, $s_i$ can be an arbitrarily ``short" (we put this term in inverted
commas because there is no background metric available with respect to which
we could measure the length of $s_i$) segment of $c$ starting at one and 
only one vertex $v$ of $\gamma$.\\ 
Let us first recall the notion of a spin-network state \cite{16,17,18}, we 
use the notation of \cite{18} : Label the edges $e$ of a graph $\gamma$ with
nontrivial irreducible representations of $SU(2)$, that is, assign to 
each of them a spin quantum number $j_e>0$. If $\pi_j$ is an irreducible
representation of $SU(2)$ with weight (or spin) $j$ then the spin-network
state depends on $\pi_{j_e}(h_e)$. Further, assign to each vertex
$v$ of $\gamma$ a contractor matrix $m_v$ which contracts the matrix elements
of the tensor product (over the set of edges $e$ incident at $v$) of the 
$\pi_{j_e}(h_e)$ in such a way that the resulting state is gauge invariant.
We label these spin-network states by $T_{\gamma,\vec{j},\vec{m}}$. One 
can show that spin-network states form an orthonormal basis of $\cal H$.\\ 
Consider first the case that $\gamma$ and $s:=s_i$ intersect in
only one point. Since $\hat{l}_s^2$ is gauge-invariant, the result of
applying $\hat{l}_s^2$ to a gauge invariant cylindrical function must be 
a gauge invariant cylindrical function $f'$ which depends on the graph
$s\cup\gamma$. Let us decompose $f'$ into a basis of spin-network states.
Since $f$ did not depend on $s$, the spin assigned to
$s$ for a particular term $T$ in the spin-network decomposition of $f'$ can 
only be $j=\pm 1,0$. In the case $j=0$ the state $T$ does not depend on
$s$ at all. In case that $j=\pm 1$ then $T$ would be a spin-network state
which contains an univalent vertex, namely the endpoint of $s$ distinct 
from $v$, and there is only one edge, namely $s$, incident at it and 
coloured with $j\not=0$. Such a state is not gauge-invariant. Therefore
the case $j\not=0$ does not appear.\\
Now consider the case that $s:=s_i$ is contained in an edge $e$ of 
$\gamma$ and
starts at a vertex $v$ of $\gamma$. By choosing $n$ high enough we 
may assume that $s$ is properly contained in $e$, that is, 
$e':=e-s\not=\emptyset$. Without 
loss of generality we may assume that $f$ is a spin-network state and 
thus $e$ carries spin $j>0$. Then in the spin-network decomposition
of $f'$ a particular term depends on $\gamma=\gamma\cup s$ in which the 
spins 
of all the edges of $\gamma$ distinct from $e$ are unchanged while $e'$
carries spin $j$ and $s$ carries spin $j'\in\{j,j\pm 1\}$. Consider the 
divalent ``vertex" $p=s\cap e'$ of $\gamma$. In case that
$j'=j$ then $T$ depends on all of $e$ with the same spin, i.e. it does not 
depend on $s$ and $e'$ with different spins. The only gauge invariant
way how to contract the corresponding matrix elements at $p$ is such that
the resulting state depends on $s,e'$ only through their product 
$e=s\circ e'$. In case that
$j'=j\pm 1$ then $p$ is a two-valent vertex at which two edges
$e',s$ with distinct spins $j,j'=j\pm 1$ are incident. There exists no
vertex contractor that makes such a state 
gauge invariant, therefore this case cannot appear.\\
This furnishes the proof that for a spin-network state $f$ 
the function $\hat{L}_n(c)f$ still depends on all of
the edges of $\gamma$ if $f$ did and that there is no dependence on the
segments of the partition of $c$ which is to be expected since the 
classical length
of a curve is a functional of $E^a_i$ only. Thus, the limit $n\to\infty$
is already taken in (\ref{13}).

\subsection{Cylindrical consistency}

Since we have adapted our partition to the graph on which a cylindrical 
function depends, although the operator (\ref{13}) was derived, it is
not obvious any more that the family of operators $(\hat{L}_\gamma(c))$
constructed actually line up and qualify as the projections of an
operator on $\cal H$. So let $\gamma\subset\gamma'$. We need to show that
1) $(\hat{L}_{\gamma'}(c))_{|\gamma}=\hat{L}_\gamma(c)$, that is, the 
restriction of a projection to a smaller graph actually coincides with 
the projection to the smaller graph and 2) the domain of 
$\hat{L}_\gamma(c)$ is contained in that of $L_{\gamma'}(c)$. 
Issue 2) follows trivially by inspection if we choose the domain of
$\hat{L}_\gamma(c)$ to be $\mbox{Cyl}_\gamma^3(\agb)$ which we can since 
the volume operator has that domain. Issue 1) follows immediately :
Given a curve $c$, the action of $\hat{L}_\gamma(c)$ and 
$\hat{L}_{\gamma'}(c)$
on a function $f$ cylindrical with respect to $\gamma$ differ if either
a) and $\gamma'$ has more vertices intersecting $c$ than $\gamma$ or b)
there are additional edges of $\gamma'$ incident at a common vertex of
$\gamma$ and $\gamma'$ intersecting $c$. This follows directly by 
inspection from (\ref{13}). So let first $v$ be a vertex of $\gamma'$ but 
not of $\gamma$ and $s$ a segment of $c$ incident of $v$. Then 
$[h_s^{-1},\hat{V}_v]f=-f\hat{V}_vh_s^{-1}=0$ and so this term in the sum
(\ref{13}) does not contribute. Now if case b) occurs then the 
cylindrical consistency of the volume operator assures that the edges of
$\gamma'-\gamma$ incident at $v$ do not contribute.\\
This furnishes the proof that (\ref{13}) is consistently defined.

\subsection{Symmetry and positivity}

We wish to show that 1) every projection $\hat{L}_\gamma(c)$ is a 
symmetric operator on $\cal H$ with domain $\mbox{Cyl}_\gamma^3(\agb)$
and 2) the family of projections $(\hat{L}_\gamma(c))_\gamma)$ comes 
from a positive semi-definite symmetric operator on $\cal H$ with domain 
$\mbox{Cyl}^3(\agb)$.
\begin{Lemma}
Every projection $\hat{L}_\gamma(c)$ defines a symmetric and positive
semi-definite operator on $\cal H$ with domain 
${\cal D}_\gamma:=\mbox{Cyl}_\gamma^3 (\agb)$.
\end{Lemma}
Proof :\\
Let us begin with $\hat{l}_s^2$. Since $\hat{V}_v$ is a symmetric operator
on $\cal H$ we find for the adjoint of $[h_s^{-1},\hat{V}_v]$ using the 
unitarity of $SU(2)$ 
\be \label{14}
[h_s,\hat{V}_v]^\dagger=[\hat{V}_v,\overline{h_s}]=-[(h_s^{-1})^T,\hat{V}_v]
\ee
where $(.)^T$ denotes the transpose of $(.)$. It follows that
\be \label{15}
\hat{l}_s^2:=+8\sum_{A,B} 
\hat{K}_{AB}(s)^\dagger\hat{K}_{AB}(s)\mbox{ where }
\hat{K}_{AB}(s)=[(h_s^{-1})_{AB},\hat{V}]
\ee
which shows that $\hat{l}_s^2$ is a symmetric and positive semidefinite
operator on $\cal H$ and therefore possesses a square root $\hat{l}_s$.
The statement about the domain is a consequence of the fact that 
$\hat{L}_\gamma(c)$ is defined in terms of the spectral projections
of $\hat{V}_\gamma$ which has domain ${\cal D}_\gamma$.\\
$\Box$\\
Let us derive a more convenient expression for (\ref{13}) :\\
By means of the basic $SL(2,\Co)$ formula 
$g^{-1}_{AB}=\epsilon_{AC}\epsilon_{BD} g_{DC}$ we readily derive that
$\hat{l}_s=\hat{l}_{s^{-1}}$. This motivates the following notation :\\
Let be given an oriented curve $c$ and a vertex $v$ of $\gamma$ that 
intersects it. If $v$ is an interior point of $c$ consider any two
segments $s_v^\pm$ of $c$ which start at $v$ and are ``short" enough as 
not to intersect $\gamma$ in any other of its vertices. If $v$ is an 
endpoint of $c$ we need only one such segment $s_v^+=s_v^-$ starting at $v$.
In (\ref{13}) the segments $s_i$ of $c$ starting at vertices of 
$\gamma$ come with the orientation of $c$, however, as we just showed, the 
operator $\hat{l}_s$ is invariant
under reversal of the orientation of $s$ and the result of applying it
to a gauge invariant function is independent of the choice of $s$ as long
as it is ``short" enough. 
Therefore, let 
\be \label{16}
N(v,c)=\left\{ \begin{array}{ll}
0 & v\;\not\in\;c\\
\frac{1}{2} & v\mbox{ is an endpoint of }c\\
1 & v\mbox{ is an interiour point of }c
\end{array} \right.
\ee
and define $\hat{L}^\pm_v(c):=\hat{l}_{s_v^\pm}(c)$. Then the length operator
can be conveniently written
\be \label{17}
\hat{L}_\gamma(c)=\sum_{v\in V(\gamma)} N(v,c)
[\hat{L}^+_v(c)+\hat{L}^-_v(c)] \;.
\ee
\begin{Theorem}
The family $(\hat{L}_\gamma(c),{\cal D}_\gamma)$ defines a positive 
semi-definite symmetric operator on ${\cal D}:=\mbox{Cyl}^3(\agb)$.
\end{Theorem}
Proof :\\
So far we have demonstrated $\hat{L}_\gamma(c)$ is a symmetric and positive 
semi-definite operator for each $\gamma$ with domain 
${\cal D}_\gamma=\mbox{Cyl}_\gamma^3(\agb)$. But to qualify 
as a symmetric projection of a symmetric operator requires more : \\
Let $\hat{S}$ be a symmetric operator on $\cal H$ with symmetric projections
$\hat{S}_\gamma$. Let $f=p_\gamma^\ast f_\gamma$ and 
$g=p_{\gamma'}^\ast g_{\gamma'}$ be cylindrical functions. Then
\be \label{18}
<g,\hat{S} f>=<g,\hat{S}_\gamma f>=<\hat{S}_\gamma g,f>\stackrel{!}{=}
<\hat{S}g,f>=<\hat{S}_{\gamma'}g,f> \;.
\ee
Condition (\ref{18}) is necessary and sufficient for a family of 
self-consistent symmetric projections
to come from a symmetric operator. Let us check that (\ref{18}) is satisfied
for $\hat{L}(c)$ for each curve $c$.\\
It is sufficient to check it for the case that $f,g$ are spin-network states.
Now since $\hat{L}_\gamma(c)$ does not change the graph $\gamma$ on which
$f$ depends non-trivially, it follows that $<g,\hat{L}_\gamma(c)f>$ is
non-vanishing if and only if $\gamma=\gamma'$ because spin-network states
depending on different graphs are orthogonal \cite{16,17,18}. 
The same holds for $<\hat{L}_{\gamma'}(c)g,f>$. Symmetry now follows
from the cylindrical consistency and the statement about the domain is 
because $\hat{V}$ has this domain.\\
$\Box$

\subsection{Self-adjointness}

Although the expression of $\hat{L}_\gamma(c)f$ does not depend on the 
$s_v^\pm$ such that there must be a way to write purely in terms of 
the right invariant vector fields $X_I$ very much like as in the case of the 
volume operator (\ref{6}), it is not clear to us how to do that. 
We can therefore not employ the essential self-adjointness
of $iX_I$ with core $\mbox{Cyl}^1(\agb)$ in order to show that 
$\hat{L}_\gamma(c)$ is 
essentially self-adjoint as well with core $\mbox{Cyl}^3(\agb)$. \\
An independent proof which demonstrates that there exist self-adjoint 
extensions goes as follows : 
\begin{Theorem}
The operator $(\hat{L}(c),{\cal D})$ admits self-adjoint extensions on 
$\cal H$. \end{Theorem}
Proof :\\
The expression of $\hat{L}(c)$
is purely real and symmetric. Therefore it commutes with the antiunitary 
operator of complex conjugation. The assertion now follows from von 
Neumann's theorem \cite{19}, p. 143.\\
$\Box$\\
A possible choice of extension is given by its Friedrichs 
extension. \\

\section{Spectral analysis of the length operator}

\subsection{Discreteness}

The important feature of $\hat{L}_\gamma(c)$ is that it changes neither
the graph $\gamma$ on which a cylindrical function depends nor the 
irreducible representations of the corresponding spin-network states into
which a cylindrical function $f$ can be decomposed. Since the number
of spin-network states associated with a fixed graph and a fixed 
colouring of its edges by irreducible representations is finite-dimensional
\cite{18} it follows that the length operator on each such subspace of 
$\cal D$ is just a symmetric, positive semi-definite, finite 
dimensional matrix. Its
eigenvalues are real and are given by certain non-negative functions 
$\lambda(\{j_I\})$ of the spins $j_I$ associated with the edges of $\gamma$. 
Since spin-networks span ${\cal D}_\gamma$ and provide a countable basis 
for its completion ${\cal H}_\gamma$ (with respect to $\mu_{0,\gamma}$)
it follows that there exists a 
countable basis of eigenvectors for ${\cal H}_\gamma$. The 
corresponding eigenvalues therefore form a countable set and lie in 
the point spectrum of $\hat{L}_\gamma(c)$.\\
Recall that a point in the spectrum is said to be in the discrete spectrum if 
it is an isolated point and an eigenvalue of finite multiplicity 
(clearly $0$ has infinite multiplicity, all functions cylindrical with 
respect to graphs not intersecting the curve $c$ are annihilated). In 
order to show that the point spectrum that $\hat{L}_\gamma(c)$ attains on
${\cal D}_\gamma$ is discrete it would be sufficient to show that
$\lambda(\{j_I\})$ diverges whenever $j:=j_1+..+j_n$, $n$ being the 
number of edges of $\gamma$, diverges. Namely,
if there was a finite condensation point in the point spectrum then an 
infinite number of different $n$-tuples would give an eigenvalue which
lies in a finite neighbourhood of that condensation point but necessarily 
the corresponding values of $j$ must diverge which would be a contradiction.
The same argument shows that there would not be an eigenvalue of infinite
multiplicity. \\
Looking at the eigenvalue 
formulae derived below for some simple graphs, such a behaviour of 
$\lambda$ is 
plausible because $\hat{L}_\gamma(c)$ is an unbounded operator, however,
a strict proof is missing at this point. \\
Discreteness of the spectrum
of $\hat{L}_\gamma(c)$ would follow immediately if we could prove it for 
the volume operator. An elegant method of proof would be to show that
$\hat{V}$ is (a positive root of) an elliptic operator. Since this operator
acts on the compact manifold $SU(2)^n$ standard results form harmonic 
analysis would imply that its spectrum is discrete. Unfortunately this does
not work : the volume operator is easily seen to be the fourth root of
the operator $\hat{q}^\dagger\hat{q},\;\hat{q}=\sum_{IJK} 
\epsilon_{IJK}\hat{q}_{IJK}$ (compare \cite{6}) and so is a 6-th order
homogenous polynomial in the derivative operators $X_I$. Its principal
symbol can be seen to be a non-negative function with an at least 
two-dimensional kernel and therefore is far from being invertible. Indeed,
it is by now well-known that the volume operator has a kernel which
includes all divalent and trivalent graphs.\\
Summarizing, we have shown that $\hat{L}_\gamma(c)$ has pure point 
spectrum when restricted to ${\cal D}_\gamma$. Now consider
the complete spectrum of $\hat{L}_\gamma(c)$ on all of ${\cal H}_\gamma$ 
Since $\hat{L}_\gamma(c)$
is a self-adjoint (not only symmetric) operator, it has spectral 
projections which are orthogonal for disjoint subsets of its spectrum.
It follows that if $I$ is a subset of the real numbers not contained in the 
part of
the spectrum that $\hat{L}_\gamma(c)$ attains on ${\cal D}_\gamma$ and if 
$\hat{P}_I$ is the
corresponding spectral projection then ${\cal H}_\gamma':=\hat{P}_I{\cal 
H}_\gamma$ 
and ${\cal D}_\gamma$ are orthogonal. But ${\cal D}_\gamma$ is dense in 
${\cal H}_\gamma$, therefore for
each $v\in{\cal H}_\gamma',\;\epsilon>0$ we find $f\in{\cal D}_\gamma$ such 
that $||v-f||<\epsilon$. On the other hand by orthogonality 
$||v-f||\ge||v||$ which is a contradiction unless $v=0$. Therefore
the complete spectrum of $\hat{L}_\gamma(c)$ is already attained on 
${\cal D}_\gamma$ and it has no continuous part very much 
like its companions, the area and volume operators \cite{10,11}. 
Thus, the spectrum on ${\cal H}_\gamma$ is discrete in the sense that it 
does not have a continuous part.

Similarly, it follows that the spectrum of $\hat{L}(c)$ is attained on 
$\cal D$ and also does not have a continuous part (this property is 
shared by all three operators, length, area and volume) : Although the 
set of 
piecewise analytic graphs is uncountable, the matrix elements of the 
length operator in a spin-network basis do not depend on the graph, they
only depend on the quantum numbers $\vec{j},\vec{m}$ {\em and are 
therefore diffeomorphism invariant}. Therefore, the spectrum attained 
on ${\cal D}_\gamma$ and ${\cal 
D}_{\varphi(\gamma)},\;\varphi\in\mbox{Diff}(\Sigma)$ an analyticity 
preserving diffeomorphism, is identical. Moreover, the spectrum does not 
depend on whether two curves touch each other in a $C^m$ or $C^n$ fashion
for any $1\le m<n$. It follows that the spectrum only depends on the 
$C^\infty$ properties of the diffeomorphism class of a graph, that is, all 
that matters is whether two edges are distinct or coincide. In other words
all we need to know about a graph is \\
a) the number $N$ of its vertices\\
b) the valence $n(v)$ of each vertex $v$\\
c) the topologically different ways of connecting edges incident at 
different vertices in a $C^\infty$ manner which is a finite number.\\
Thus, since this characterization of a graph depends on discrete labels 
and the union of countable sets is a countable set it follows that 
the spectrum is still discrete in the sense that it does not have a 
continuous part. On the other hand we see that every eigenvalue of the 
length, area or volume operators is of uncountably infinite multiplicity 
when we consider all of $\cal H$.

\subsection{Eigenvalues}

By inspection, the task of giving a closed formula for the eigenvalues 
of the length operator requires to have a closed formula for the spectrum of
the volume operator at one's disposal which we lack, however, at the present
stage (see, however, \cite{22} for a closed formula for its matrix 
elements). We will therefore restrict ourselves here to compute the 
spectrum for some simple types of graphs and thereby obtain the quantum
of length.\\
Specifically, it is comparatively simple to compute the spectrum of the 
length operator when restricted to at most trivalent graphs, thus including
the classical spin-networks which were introduced in \cite{Penrose} and whose
vertices are all precisely trivalent. Since for trivalent graphs the 
vertex contractors of a spin-network are unique and since the result
of applying the length operator to a spin-network state is a spin-network
state on the same graph and with the same spin, it follows that all
spin-network states on trivalent graphs are eigenvectors of the 
length operator. We will see that the corresponding eigenvalues are 
non-vanishing in general. This is an astonishing feature because the 
volume operator is known to vanish on trivalent graphs. Now, classically
the volume of a region is known if one can measure the length of 
arbitrary curves through that region and so non-vanishing length of 
curves results in non-vanishing volume. This indicates that trivalent
graphs are rather special and are insufficient to construct weave states
that approximate a given classical geometry.\\ 
The computations are largely governed by the properties of the 
$6j-$symbol of the recoupling theory of angular momentum which we recall
in the appendix. The way how the recoupling theory of spin systems 
enters the stage is as follows : in our computations we evaluate the volume
operator at a given vertex $v$ on functions $f$ which transform according
to an irreducible representation $j$ of $SU(2)$ under gauge transformations 
at $v$. This $j$ is nothing else than the resulting total angular 
momentum to which
the angular momenta $j_I$ (which colour the edges $e_I$ incident at $v$)
couple. There are different ways of how to couple angular momenta $j_I$
to resulting spin $j$ and this freedom is determined by the recoupling
quantum numbers $j_{IJ}, j_{IJK},..$ (compare the appendix) and results in 
different
contractors of the so not necessarily gauge invariant (or extended)
spin networks. Here we mean by an extended spin network just any function
that depends on the edges of a graph through the matrix elements 
of irreducible representations of $SU(2)$ evaluated at the holonomy of 
the corresponding edge and transforms at each vertex according to an
irreducible representation of $SU(2)$ under gauge transformations. The 
point is now that the function $f$ is easily seen to
be in the left regular representation of $SU(2)^n$, $n$ being the valence 
of the vertex $v$, defined by $\hat{R}(g_1,..,g_n)f(h_1,..,h_n)=
f(g_1 h_1,..,g_n h_n)$. On the other hand, the connection with the abstract
angular momentum Hilbert space where $SU(2)$ acts by the abstract unitary 
representation $\hat{U}(g)$ and which is spanned by the vectors of the form 
$|(j_1,m_1),..,(j_n,m_m);k_1,..,k_n>
=|j_1,m_1;k_1>\otimes..\otimes|j_n,m_n;k_n>$ 
where $k_I$ are certain additional quantum numbers, is 
made as follows : Consider the special functions 
$(\pi_j(h))_{m,m'}$ given by the matrix elements of the $j-th$ irreducible
representation of $SU(2)$, $m,m'\in\{-j,-j+1,..,j\}$. Consider the 
Hilbert space $L_2(SU(2),d\mu_H)$ and let $|g>$ be the usual Dirac
generalized eigenstates of the multiplication operator $\hat{g}$. Then
$(\pi_j(h))_{m,m'}=<h|j,m;m'>$. The proof is by checking the representation
property $\hat{R}(g)(\pi_j(h))_{m,m'}=(\pi_j(gh))_{m,m'}=
(\pi_j(g))_{m,\tilde{m}}(\pi_j(h))_{\tilde{m},m'}
=<h|\hat{U}(g)|j,m;m'>$ by definition of the states $|j,m;m'>$.
The reader is referred to \cite{22} for more details.\\
It follows from these considerations that instead of doing tedious 
computations within the left regular representation (spin-network
states or traces of the holonomy around closed loops, \cite{21}) it is far 
easier to do them in the abstract representation
which allows to use the powerful Clebsh-Gordan theory of angular momentum.
We will do that in the sequel.\\
In particular, it follows immediately that the right invariant vector
field $X_I$ is identified with $2i J_I$ where $J_I$ is the angular 
momentum operator of the spin associated with the $I-th$ edge which is the
self-adjoint generator of the unitary group $\hat{U}(g)$.\\
We are now in the position to compute the spectrum of the length
operator on trivalent graphs $\gamma$. 
First we cast the volume operator in the more convenient form
\ba \label{19}
\hat{V} &=& \ell_p^3 \hat{v}f=\ell_p^3 \sum_{v\in V(\gamma)}\hat{V}_v 
\mbox{ where } \hat{V}_v:=\sqrt{|\frac{i}{32}\hat{q}_v|}\nonumber\\
\hat{q}_v&=&\sum_{e_I\cup e_J\cup e_K=v,I<J<K} 
\epsilon(e_I,e_J,e_K)\hat{q}_{IJK}\mbox{ where }
\hat{q}_{IJK}:=[J_{IJ}^2,J_{JK}^2]\;.
\ea
To arrive at this expression one only has to use elementary angular momentum 
algebra. Expression (\ref{19}) captures a neat interpretation of the 
volume operator : it measures the difference between recoupling schemes
of $n$ angular momenta based on $J_{IJ}$ and $J_{JK}$ respectively.
This is why the recoupling theory is important and the matrix elements 
of the volume operator can be given purely in terms of polynomials of
$6j-$symbols \cite{22}.\\
Now let $v$ be a vertex of $\gamma$
which intersects $c$. From (\ref{17}) we find 
($h:=h_{s_v^\pm},\hat{L}_v(c):=\hat{L}_v^\pm(c)$) 
\be \label{20}
\frac{1}{8}\hat{L}_v(c)^2=-[\mbox{tr}(h^{-1}\hat{V}h)\hat{V}+
\hat{V}\mbox{tr}(h^{-1}\hat{V}h)]+2\hat{V}^2+\mbox{tr}(h^{-1}\hat{V}^2h)
\ee
and the simplification that occurs on trivalent graphs is that the first
three terms vanish identically.
We have two possibilities :\\
Case A : $s_v$ lies within an edge of $\gamma$ incident at $v$ or\\
Case B : $s_v$ is not contained in an edge of $\gamma$.\\
We will discuss both cases separately. 

\subsubsection{Case A }

We may without loss of generality assume that $s_v$ coincides with
one, say $e_3$, of the three edges of $\gamma$ incident at $v$.
This is because 1) if $e=s_v\circ e'$ then for any irreducible representation
$\pi_j$ of $SU(2)$ we have $\pi_j(h_e)=\pi_j(h_{s_v})\pi_j(h_{e'})$ so
a spin network state also depends on $s_v$ through $\pi_j$ and 2) for a right
invariant vector field $X(h_{s_v})=X(h_{s_v} h_{e'})=X(h_e)$ by definition
of right invariance. Therefore the volume operator (\ref{19}) contains only
one term $\hat{q}_{123}$. \\
Consider a trivalent spin-network function $f=T_{\gamma,\vec{j}}$ with 
spins $j_1,j_2,j_3$ assigned to $e_1,e_2,e_3$ (we have suppressed the 
contractor matrices $\vec{m}$ since they are unique for trivalent 
spin-networks). Then $h_{e_3}f$ can be decomposed into extended spin network 
functions with total angular momentum $j=1/2$ and edge spins 
$j_1'=j_1,j_2'=j_2,j_3'=j_3\pm 1/2$ by usual Clebsh-Gordan representation
theory. Let us determine the precise coefficients of that decomposition.
\begin{Lemma} \label{la1}
Let $n=2j$ so that $\pi_j(g)_{A_1,..,A_n;B_1,..,B_n}=
g_{(A_1,B_1}..g_{A_n),B_n}$ where the round bracket means total 
symmetrization in the $A$ indices. Then
\ba \label{21}
g_{A_0,B_0}\pi_j(g)_{A_1,..,A_n;B_1,..,B_n}&=&
\pi_{j+1/2}(g)_{A_0,..,A_n;B_0,..,B_n}\nonumber\\
&& -\frac{n}{n+1}\epsilon_{A_0(A_1}\pi_{j-1/2}(g)_{A_2,..,A_n);(B_2,..,B_n}
\epsilon_{B_1)B_0}\;.
\ea
\end{Lemma}
Proof :\\
Elementary linear algebra.\\
$\Box$\\
We now multiply each of the two terms in (\ref{21}) with $g^{-1}_{B_0,A_0}$
and sum over $A_0,B_0$. The result is 
\ba \label{22} 
&& g^{-1}_{B_0,A_0}\pi_{j+1/2}(g)_{A_0,..,A_n;B_0,..,B_n}=
\frac{n+2}{n+1}\pi_j(g)_{A_1,..,A_n;B_1,..,B_n} \nonumber\\
&&g^{-1}_{B_0,A_0}\epsilon_{A_0(A_1}\pi_{j-1/2}(g)_{A_2,..,A_n);(B_2,..,B_n}
\epsilon_{B_1)B_0}=-\pi_j(g)_{A_1,..,A_n;B_1,..,B_n}\;.
\ea
Formulae (\ref{22}) illustrate the computational reason for why the edge
$s_v$ does not appear in a gauge invariant function $f$ after evaluating
$\hat{L}(c)f$ on it. We now can write $h_{e_3}f=f_++\frac{n_3}{n_3+1}f_-$
where, according to (\ref{21}), the vectors $f_+$ and $f_-$ respectively are
proportional to the vectors $|j_{12}=j_3,j=1/2;j_1,j_2,j_3+1/2>$ and
$|j_{12}=j_3,j=1/2;j_1,j_2,j_3-1/2>$ respectively on the abstract angular
momentum Hilbert space. That in both vectors $j_{12}$ still equals $j_3$
follows from the fact that we did not change the way the matrices are 
contracted in $f$ in multiplying it by $h_{e_3}$.\\
The space of states $V_+$ with total angular momentum $j=1/2$ and spins
$j_1,j_2,j_3'=j_3+1/2$ is two dimensional : it is spanned by
$|j_{12}=j_3,j=1/2;j_1,j_2,j_3+1/2>$ and 
$|j_{12}=j_3+1,j=1/2;j_1,j_2,j_3+1/2>$ respectively. Likewise, the span
of the space of states $V_-$ with total angular momentum $j=1/2$ and spins
$j_1,j_2,j_3'=j_3-1/2$ is given by $|j_{12}=j_3,j=1/2;j_1,j_2,j_3-1/2>$ and,
provided that $j_3\ge 1$, $|j_{12}=j_3-1,j=1/2;j_1,j_2,j_3-1/2>$ 
respectively. The volume operator leaves these two spaces separately
invariant and the operator $\hat{q}_{123}$, being antisymmetric on $\cal H$, 
reduces to an antisymmetric 
$2\times 2$ matrix on $V_+$ and to an anti-symmetric $2\times2$ matrix
on $V_-$ if $j_3\ge1$ and to the zero matrix if $j_3<1$. Let the matrix
elements different from zero of these matrices be denoted by $\pm\mu_+,
\pm\mu_-$ respectively,
then, since we take the modulus of the square root of $\hat{q}_{123}$ it 
follows that $f_\pm$ are already eigenvectors of $\hat{V}$ with eigenvalues
$\lambda_\pm:=\frac{\sqrt{|\mu_\pm|}}{4\sqrt{2}}$. It follows then from 
(\ref{22}) and $h^{-1}\hat{V}^2h=(h^{-1}\hat{V}h)^2$ 
that our spin-network state is eigenfunction of the 
length operator $\hat{L}_v(c)$ with eigenvalue 
$\frac{\ell_p}{2}\sqrt{\frac{1}{n_3+1}[(n_3+2)\lambda_+^2+n_3\lambda_-^2]}$.\\
It remains to compute the matrix elements $\mu_\pm$ of 
$\hat{q}_{123}=[J_{12}^2,J_{23}^2]$.
We have 
\ba \label{23}
 \mu_+ &:=& <j_{12}'=j_3+1,j=\frac{1}{2};j_1,j_2,j_3'=j_3+\frac{1}{2}|
\hat{q}_{123}|j_{12}=j_3,j=\frac{1}{2};j_1,j_2,j_3'=j_3+\frac{1}{2}> 
\nonumber\\ &=& [(j_3+1)(j_3+2)-j_3(j_3+1)]
<j_{12}'=j_3+1,j=\frac{1}{2}|J_{23}^2|j_{12}=j_3,j=\frac{1}{2}>
\nonumber\\
&=& 2(j_3+1)\sum_{j_{23}=j_1\pm\frac{1}{2}} j_{23}(j_{23}+1)
<j_{12}'=j_3+1,j=\frac{1}{2}|j_{23},j=\frac{1}{2}>\times\nonumber\\
&& \times <j_{23},j=\frac{1}{2}|j_{12}=j_3,j=\frac{1}{2}>
\ea
where we have suppressed $j_1,j_2,j_3'=j_3+\frac{1}{2}$. In the last step we 
have inserted a complete $1$ in form of the coupling scheme $(j_2,j_3')\to 
j_{23},\;(j_1,j_{23})\to j$ and realized that the only possible values
for $j_{23}$ in order to couple to spin $j=1/2$ with $j_1$ are the ones 
displayed. Completely analogously we find
\ba \label{24}
 \mu_- &:=& <j_{12}'=j_3-1,j=\frac{1}{2};j_1,j_2,j_3'=j_3-\frac{1}{2}|
\hat{q}_{123}|j_12=j_3,j=\frac{1}{2};j_1,j_2,j_3'=j_3-\frac{1}{2}>\nonumber\\
&=& [(j_3-1)j_3-j_3(j_3+1)]
<j_{12}'=j_3-1,j=\frac{1}{2}|J_{23}^2|j_{12}=j_3,j=\frac{1}{2}>
\nonumber\\
&=& -2j_3\sum_{j_{23}=j_1\pm\frac{1}{2}} j_{23}(j_{23}+1)
<j_{12}'=j_3-1,j=\frac{1}{2}|j_{23},j=\frac{1}{2}>\times\nonumber\\
&& \times <j_{23},j=\frac{1}{2}|j_{12}=j_3,j=\frac{1}{2}>\;.
\ea
We now use the formulae given in the appendix to compute the eight
remaining matrix elements in terms of $6j-$symbols and evaluate
the latter by means of the Racah formula. The result is
\ba \label{25}
&&\mu_+=\sqrt{(a+b+c)(a+b-c)(b+c-a)(c+a-b)}\nonumber\\
&& \mbox{ where }a=j_1+\frac{1}{2},b=j_2+\frac{1}{2},
c=j_3+1\nonumber\\
&&\mu_-=-\sqrt{(a+b+c)(a+b-c)(b+c-a)(c+a-b)}\nonumber\\
&& \mbox{ where }a=j_1+\frac{1}{2},b=j_2+\frac{1}{2},
c=j_3 \;.
\ea
The final result is thus given by the following theorem.
\begin{Theorem} \label{th1}
The eigenvalue $\lambda$ of $\hat{L}_v(c)$ for a trivalent spin-network 
state $T_\gamma$ with vertex at $v$ such that $c$ and $\gamma$ share a 
segment incident at $v$ is 
given by ($j_3\in\{j_1+j_2,j_1+j_2-1,..,|j_1-j_2|\}$) 
\ba \label{26}
&&\lambda=\frac{\ell_p}{2\sqrt{j_3+1/2}}\times\nonumber\\
&& \times \sqrt{(j_3+1)\sqrt{(j_1+j_2+j_3+2)(j_1+j_2-j_3)(j_2+j_3-j_1+1)
(j_3+j_1-j_2+1)}}
\nonumber\\
&& 
\overline{+j_3\sqrt{(j_1+j_2+j_3+1)(j_1+j_2-j_3+1)(j_2+j_3-j_1)(j_3+j_1-j_2)}}
\ea
provided the edges $e_1,e_2,e_3$ of $\gamma$ incident at $v$ are linearly
independent, otherwise $\lambda=0$.
\end{Theorem}
The formula for $\lambda$ simplifies if we take $v$ to be a divalent
vertex not sharing a segment with $c$. Then $j_3=0$ and necessarily 
$j_1=j_2=j_0$ 
and we find simply \[\lambda(j_0)=\ell_p\root 4\of {j_0(j_0+1)}\]
which for large spin grows as $\sqrt{j_0}$. In 
other words, the warping
of $\Sigma$ when the gravitational field is in the spin-network state 
labelled by $j_0$ grows with the spin $j_0$ since one and the same curve 
$c$ appears the longer the more spin $j_0$ we have in a neighbourhood 
of it. \\
The quantum of length is 
achieved for $j_0=1/2$ and given by 
$$
\frac{1}{\sqrt{2}}\root 4\of{3}\ell_p. 
$$
The length can change only in packets of $\Delta L\approx\pm 
\frac{1}{2}\sqrt{\frac{1}{j_0}}\ell_p$ which
approaches zero for large spin so that for high spin the length looks
like a continuous operator. 

\subsubsection{Case B}

As this is actually a four-valent problem the computations will be much
more elaborate than in the previous case since we have to couple 
now four angular momenta which will necessarily invoke the $9j-$symbol.
Fortunately one can reduce the $3(n-1)j-$symbol to a polynomial in
$6-j$symbols, $n$ the valence of the vertex.\\
The segment $e_4:=s_v$ carries
spin $j_4=1/2$. First we note that if again $f$ is a spin-network state
with vertex $v$ intersecting $c$ then we have ($\epsilon_{IJK}:=
\epsilon(e_I,e_J,e_K)$)
\ba  \label{27}
\hat{q}_v h_{e_4}f &=&
[\epsilon_{124}\hat{q}_{124}+\epsilon_{234}\hat{q}_{234}+
\epsilon_{314}\hat{q}_{314}]h_{e_4}f+h_{e_4}\epsilon_{123}\hat{q}_{123}f
\nonumber\\
&=&[\epsilon_{124}+\epsilon_{234}+\epsilon_{124}]\hat{q}_{124}h_{e_4}f
\ea
where in the second step we exploited gauge invariance of $f$, that is,
in infinitesimal form $[X_1+X_2+X_3]f=0$.\\
The extended spin-network function $h_{e_4}f$ is represented on the abstract
angular momentum Hilbert space by a vector proportional to 
$\psi:=|j_{12}=j_3,j_{123}=0,j=1/2;j_1,j_2,j_3,j_4=1/2>$ where the 
notation 
means that the operator $[(J_1+J_2+J_3)^i]^2$ is diagonal as well, the
coupling scheme being given by $(j_1,j_2)\to j_{12}, (j_{12},j_3)\to j_{123},
(j_{123},j_4)\to j$.\\
The space of states with given values $j_1,j_2,j_3,j_4=j=1/2$ is easily
seen to be three-dimensional and is spanned by $\psi,\psi_+,\psi_-$
where $\psi_\pm=|j_{12}=j_3\pm 1,j_{123}=1,j=1/2;j_1,j_2,j_3,j_4=1/2>$.
This is because $j=1/2$ requires $j_{123}=0,1$ and $j_{123}=0$ enforces
$j_{12}=j_3$ while $j_{123}=1$ enforces $j_{12}=j_3\pm 1$ by the usual
Clebsh-Gordan decomposition into irreducibles 
$(j)\otimes(j')=(j+j')\oplus(j+j'-1)\oplus..\oplus(|j-j'|)$.
The task left to do is to compute the matrix elements of 
$q_{124}=[J_{12}^2,J_{24}^2]$ between $\psi_,\psi_\pm$. In the following
computation we are going to insert a complete $1$ in form of the
basis corresponding to the coupling scheme 
$(j_2,j_4)\to j_{24},(j_1,j_{24})\to j_{124}, (j_{124},j_3)\to j$
in order to evaluate $J_{24}^2$. We abbreviate 
$|j_{12}=j_3,j_{123}=0,j=1/2;j_1,j_2,j_3,j_4=1/2>=:|j_{12},j_{123}>$
etc. and have for any values of the various $j's$
\ba \label{28}
&& <j_{12}',j_{123}'|\hat{q}_{124}|j_{12},j_{123}>
=[j_{12}'(j_{12}'+1)-j_{12}(j_{12}+1)]
<j_{12}',j_{123}'|J_{24}^2|j_{12},j_{123}>\nonumber\\
&=&[j_{12}'(j_{12}'+1)-j_{12}(j_{12}+1)]\times\nonumber\\
&& \times\sum_{j_{24},j_{124}}
j_{24}(j_{24}+1) <j_{12}',j_{123}'|j_{24},j_{124}>
<j_{24},j_{124}|j_{12},j_{123}>
\ea
where the allowed values for $j_{24},j_{124}$ are $j_{124}=j_3\pm 1/2$
in order to couple with $j_3$ to resulting $j=1/2$ and $j_{24}=j_2\pm 1/2$
are the possible irreducible representations contained in $(j_2)\otimes
(j_4)$. The next step is to reduce the matrix elements to $6j-$symbols.
Note that the matrix elements are not of the standard textbook form
$<j_{12},j_{34}|j_{13},j_{24}>$ so that we need to invent a new
reduction of the $9j-$symbol as a product of $6j-$symbols.
Upon inserting a complete $1$ corresponding to a basis of yet another 
coupling scheme we have 
\ba \label{29}
&&<j_{12},j_{123}|j_{24},j_{124}>\nonumber\\
&\equiv&
<j_{12}(j_1,j_2),j_{123}(j_{12},j_3),j(j_{123},j_4)|
j_{24}(j_2,j_4),j_{124}(j_1,j_{24}),j(j_{124},j_3)>\nonumber\\
&=& \sum_{j_{124}'} <j_{12}(j_1,j_2),j_{123}(j_{12},j_3),j(j_{123},j_4)|
|j_{12}(j_1,j_2),j_{124}'(j_{12},j_4),j(j_{124},j_3)>\times \nonumber\\
& & \times <j_{12}(j_1,j_2),j_{124}'(j_{12},j_4),j(j_{124},j_3)|
j_{24}(j_2,j_4),j_{124}(j_1,j_{24}),j(j_{124},j_3)>\nonumber\\
&=& <j_{123}(j_{12},j_3),j(j_{123},j_4)|
|j_{124}(j_{12},j_4),j(j_{124},j_3)>\times \nonumber\\
&& \times <j_{12}(j_1,j_2),j_{124}(j_{12},j_4)|
j_{24}(j_2,j_4),j_{124}(j_1,j_{24})> \;.
\ea 
In going from the third equality to the fourth equality, two things 
happened :
a) because $j_{124}$ is diagonal on the vectors involved in the
second scalar product factor in the third line, the sum reduces to only one
term $j_{124}'=j_{124}$, b) we used the Wigner-Eckart theorem to
reduce the matrix elements \cite{20}, that is, we could get rid of the
first entry of the vectors involved in the  
first scalar product factor and of the last entry in the second, both in the 
third line. Equation (\ref{29}) is the form that 
allows to express everything in terms of $6j-$symbols. Namely we have
\ba \label{30}
&& <j_{123}(j_{12},j_3),j(j_{123},j_4)
|j_{124}(j_{12},j_4),j(j_{124},j_3)>=(-1)^{j_1+j_2+j_4+j_{124}}
\times\nonumber\\
&& \times\sqrt{(2 j_{12}+1)(2 j_{124}+1)}
\left\{ \begin{array}{ccc}
j_1 & j_2 & j_{12}\\
j_4 & j_{124} & j_{24} \end{array} \right\}
\\ \label{31}
&& <j_{12}(j_1,j_2),j_{124}(j_{12},j_4)|
j_{24}(j_2,j_4),j_{124}(j_1,j_{24})>=(-1)^{2(j_3+j_{12})+j_4+j-j_{123}}
\times\nonumber\\
&& \times\sqrt{(2j_{123}+1)(2j_{124}+1)}
\left\{ \begin{array}{ccc}
j_3 & j_{12} & j_{123}\\
j_4 & j & j_{124} \end{array} \right\} \;.
\ea
In order to see this we do the following : In (\ref{30}) set
$j_1':=j_1,j_2':=j_2,j_3':=j_4,j':=j_{124},j_{12}':=j_{12},j_{23}':=j_{24}$
and use the standard formula (\ref{a}) for the $6j-$symbol given in 
the appendix in terms of the primed $j's$.
In (\ref{31}) we first set 
$j_1':=j_3,j_2':=j_{12},j_3':=j_4,j':=j,j_{12}':=j_{123},j_{23}':=j_{124}$
and secondly recall the identity $|j_{12}(j_2,j_1),j>=(-1)^{j_1+j_2-j_{12}}
|j_{12}(j_1,j_2),j>$ in order to apply the standard formula in terms of the
primed $j's$. \\
As the whole expression becomes rather lengthy we refrain from writing
the matrix elements (\ref{28}) out explicitly in terms of $j_1,j_2,j_3$,
rather we consider them as known through (\ref{28})-(\ref{32}) and define
\be \label{32}
a:=<\psi|\hat{q}_{124}|\psi_+>,\;b:=<\psi|\hat{q}_{124}|\psi_->\;,
c:=<\psi_+|\hat{q}_{124}|\psi_->\;.
\ee
So $\hat{q}_{124}$ reduces to an antisymmetric
$3\times 3$ matrix with non-vanishing off-diagonal entries $\pm a,\pm 
b,\pm c$, eigenvalues $0,\pm i \mu$ where $\mu=\sqrt{a^2+b^2+c^2}$
and eigenvectors $\psi^0,\psi^\pm$ which can chosen to be
\be \label{33}
\psi^0=c\psi-b\psi_+ +a\psi_-,\;\psi^\pm=[\mp ib\mu-ac]\psi+[-\mp i c\mu+ab]
\psi_+ +[b^2+c^2]\psi_- 
\ee
and inverted 
\ba \label{34}
\psi_- & = & \frac{1}{2\mu^2}[\psi^+ +\psi^- +2a\psi^0], \nonumber\\
-[b^2+c^2]\psi_+ &=& \frac{b}{\mu^2}[(b^2+c^2)\psi^0-(\psi^+ +\psi^-)/2]
+\frac{ib}{2\mu}[\psi^+ -\psi^-],\nonumber\\
+[b^2+c^2]\psi &=& \frac{c}{\mu^2}[(b^2+c^2)\psi^0-(\psi^+ +\psi^-)/2]
-\frac{ic}{2\mu}[\psi^+ -\psi^-].
\ea
Let now $\sigma:=\sqrt{|\epsilon_{124}+\epsilon_{234}+\epsilon_{124}|
\frac{\mu}{32}}$,
$\sigma$ being the eigenvalue of $\hat{V}$ on both $\psi^\pm$ while it
vanishes on $\psi^0$. It is then straightforward to check that
\be \label{35}
\hat{V}\psi=\sigma[\psi-\frac{c}{\mu^2}\psi^0]=\sigma\frac{a^2+b^2}{\mu^2}
\psi+\mbox{ terms }\propto \psi_\pm\;.
\ee
We now translate this result back into the representation
$L_2(\agb,d\mu_0)$, multiply by $h_{e_4}^{-1}$ and take the trace.
The result must be a gauge invariant vector with no dependence on
$e_4$, that is, $j=j_4=0$. It follows that $\mbox{tr}(h_{e_4}\psi_\pm)=0$
because in writing $h_{e_4}\psi_\pm$ in terms of extended spin-networks 
we cannot get $j=0$ since the contractor corresponding to $j_{123}=1$
is unchanged. Therefore
\begin{Theorem} 
The eigenvalue of $\hat{L}_v(c)$ on a trivalent spin-network state 
$T_\gamma$ with vertex $v$ such that $c$ and $\gamma$ do not share a segment 
incident at $v$ is given by
\be \label{36}
\hat{L}_v(c) T_\gamma=\ell_p
\sqrt{|\epsilon_{124}+\epsilon_{234}+\epsilon_{124}|\mu}\frac{a^2+b^2}{2\mu^2}
T_\gamma\;, \mu:=\sqrt{a^2+b^2+c^2}
\ee
where $a,b,c$ are given through (\ref{28})-(\ref{32}).
\end{Theorem}

\section{Tube operator and weaves}

The length operator has a strange feature unshared by the area and volume
operators : \\
Ultimately, in a semiclassical approximation, one wishes to construct 
states which approximate a fixed classical 3-geometry $(\Sigma,q_{ab})$.
Such eigenstates have been called ``weaves" in the literature (see for 
instance \cite{24}).
Most of these eigenstates typically only involved linked, rather than 
intersecting 
loops, while the length and volume operators automatically annihilate 
those states, we need to construct more general weave states for which 
the name ``lattice states" is more appropriate.\\
It is clear that a state that approximates such a geometry has to be 
defined on an (infinite, in case $\Sigma$ is not compact) graph filling
all of $\Sigma$ in the following sense : consider a parameter $\delta$
of which we may think as a lattice spacing and which is to characterize 
the average distance (as measured by $q_{ab}$) between neighbouring 
vertices of the graph. In order to serve as a good approximation, the 
parameter needs to be small as compared to the scales of macroscopic objects.
Consider any macroscopic volume or area in $\Sigma$. It follows that the 
graph necessarily intersects these objects provided that $\delta$ is small
enough and it intersects it the more often the smaller $\delta$. Each 
intersection makes a contribution no matter whether the intersection 
point is a vertex or not. On the other hand, even for the most random 
distribution of 
vertices with mean separation $\delta$, even if $\delta$ is much smaller
than the length of a given curve, the curve genuinely almost never 
intersects the graph in a vertex and so the quantum length of the curve 
would be always much smaller than its classical length. \\
This is obviously not what we want and so the length operator constructed 
cannot be directly associated with the length of a given classical curve.
Following the old physical argument that an object always has to have a 
linear
extension of at least one Planck length in order not to form a black hole, 
the conclusion is that a one-dimensional 
curve cannot correspond to a physical object. This motivates the 
following definition.
\begin{Definition}
A tube $C$ with a curve $c$ as center corresponding to a classical 
metric $q_{ab}$ is a two parameter congruence of curves 
$C\;:\; (r,s)\in [0,1]\times[-\pi.\pi]\to C_{r,s}$ such that\\
i) $C_{0,s}=c\;\forall\;s\in[-\pi,\pi]$,\\
ii) the transversal extension of the tube as measured by
$$
\Delta:=\sup_{s,t}\int_0^1 dr\sqrt{C_{,r}^a C_{,r}^b q_{ab}}(s,t)   
$$
is much smaller than the length $L(c)$ of the central curve $c$.
\end{Definition}
So the picture is that we have a congruence of curves, all of which 
look like $c$ and which fill a cylinder with thickness $\Delta<<L(c)$.
We are now ready to define a tube operator.
\begin{Definition}
The tube operator is given by
\be \label{37}
\hat{L}(C):=\sum_{(r,s)\in[0,1]\times [-\pi,\pi]} \hat{L}(C_{r,s})\;.
\ee
\end{Definition}
Notice that there is an uncountable sum involved in (\ref{37}). In order to
see that this makes sense we first of all notice that the classical
volume of the region filled by the tube is given approximately by $\pi 
\Delta^2 L(c)$, provided that $q_{ab}$ is slowly varying there. The density
of vertices is given by $1/\delta^3$ so that we have approximately
$\pi L(c)\Delta^2/\delta^3$ vertices inside the region filled by the tube.
For a genuine distribution of vertices, each curve involved in (\ref{37})
will intersect at most one vertex. On the other hand, all these 
intersections should be assigned to the central curve only because we wish
to measure the length of the curve $c$ only. This is the reason why we do 
not divide by the number of contributing curves. \\
It follows that only a finite number of curves contribute in (\ref{37})
and so the tube operator is densely defined on $\cal D$ and it is trivial 
to see that it is positive semi-definite, symmetric and possesses 
self-adjoint extensions. Notice that since we have prescribed the tube to be
a congruence, it follows that each vertex lies on at most one curve. Since
the length operators $\hat{L}(s_v)$ commute for distinct vertices, it 
follows that all contributing length operators can be simultaneously 
diagonalized.\\
As an example assume that we wish to approximate the Euclidean metric
$q_{ab}=\delta_{ab}$ so that it is everywhere constant and not varying at 
all. For simplicity we want to consider a weave built from a tri-valent 
lattice so that all length operators are automatically diagonal. To simplify
life further, let us assume that all edges have equal spin $j_0$ which 
implies that $j_0$ is integral. Let $\lambda(j_0)$ be the corresponding 
eigenvalue in (\ref{36}) (almost never will the lattice have a segment in 
common with one of the contributing curves). Then it follows that we get
for the eigenvalue of the tube operator
\be \label{38} 
\Lambda(C)=\ell_p\frac{\pi \Delta^2 L(c)}{\delta^3}\lambda(j_0)
(1+o(\delta/L(c))
\ee
which we want to equal $L(c)$. Since we want the tube operator to be 
state-independent and since there is no other state-independent length 
scale in the problem, we are forced to choose $\Delta=\ell_p$ (which is 
also physically 
motivated as above; we have chosen a fixed factor of proportionality to 
equal unity). It then follows that $\delta=\ell_p 
\root 3 \of{\pi\lambda(j_0)}$. Since $\lambda$ is a monotonously
growing function of $j_0$ it follows that the lattice can be chosen the 
coarser the more spin it carries in order to still approximate the same 
geometry. The limit $\delta\to 0$ blows up,
it is physically meaningless to consider lattices with mean distance of 
vertices smaller than Planck scale which, as already stated in the 
literature, hints at a
discrete structure. Since we have chosen $\Delta=\ell_p\le \delta(j_0)$
it follows as a consistency check that at most one vertex per unit length
$\delta$ will contribute to $\Lambda(C)$ so that $\hat{L}(C)$ correctly
measures a one-dimensional object.\\
The limit $\delta\to\infty$ 
gives a zero eigenvalue because $\Sigma$ becomes more and more empty.\\
So we see that there are weave states 
$\Psi(q_{ab}:=\delta_{ab},j_0;\delta(j_0))$ 
which approximate $q_{ab}=\delta_{ab}$ where we have considered 
$j_0$ as a free parameter. Notice that there is a large number
of weave states corresponding to the various values of $j_0$ such that 
the same geometry is approximated (we cannot let $j_0\to\infty$ because of
$\delta<<L(c)$). Also observe that we use here weave states which 
necessarily involve intersecting and overlapping ($j_0>1/2$), rather than 
only linked, loops \cite{24}.\\
On the other hand we can fix $\delta$, vary $j_0$ and ask
which classical geometry is being approximated. Since one and the same curve
appears the longer, according to (\ref{38}), the more spin the lattice 
carries, we see that that the corresponding $q_{ab}$ must be warped as
compared to $\delta_{ab}$. This hints at the fact that the spin of the 
lattice must have something to do with the curvature which in turn has to
do with the local energy distribution of the gravitational field. Indeed, 
the spin characterizes the eigenvalue of the ADM Hamiltonian \cite{7,25}.

\section{Summary}

$\bullet$ A satisfactory quantization of the length of a piecewise smooth
curve was given. The length operator leaves the piecewise analytic graph 
of a cylindrical function invariant (in particular, analytic) therefore
it is proper to allow the curve to be only piecewise smooth rather than
piecewise analytic. It is gauge invariant and diffeomorphism covariantly
defined.\\
$\bullet$ The length operator is an unbounded, self-adjoint, positive
semi-definite operator with domain $\mbox{Cyl}^3(\agb)$. \\
$\bullet$ The spectral analysis turns out to be tedious but 
straightforward. In particular, the methods displayed here reveal that
the complete spectrum can be found by writing a suitable computer code.
This also is true for the volume operator \cite{22}.\\
$\bullet$ The spectrum is entirely discrete, the quantum of the length 
being given by $\root 4\of{3}\ell_p/$. In particular, all spin-network states
on graphs with valence not bigger than three are eigenvectors of the
length operator with, in general, non-vanishing eigenvalue.\\
$\bullet$ In order approximate a 3-geometry a tube operator has to be 
constructed. One can build intersecting and overlapping weave states which 
are such 
that they reproduce the classical length of the tubes given a classical 
geometry.
\\
\\
\\
{\large Acknowledgements}\\
\\
This research project was supported in part by DOE-Grant
DE-FG02-94ER25228 to Harvard University.

\begin{appendix}

\section{The $6j-$symbol}

The following can be found in any textbook on the recoupling theory
of angular momenta (for instance \cite{20}).\\
Consider the coupling of three angular momenta of some spin system consisting
of three independent subsystems with angular momenta $j_I,I=1,2,3$. Since
the three systems are independent of each other, the operators
$J_I^i,J_J^j,I\not=J$ mutually commute.
We are interested in states which are labelled by the quantum numbers of a 
maximal set of mutually commuting observables which includes the 
the square of the total angular momentum $J^i=J^i_1+J^i_2+J^i_3$ of this 
system. It is easy to see that the set consisting of $j_1,j_2,j_3,j,m$
(where $m$ is the eigenvalue of $J^3$) is insufficient because 
$j_i,m_i$ is a set consisting of six rather than five quantum numbers.
The missing quantum number is any choice of $j_{IJ},I<J$ which labels the 
eigenvalue of the square of $J_{IJ}:=J_I+J_J$. It is easy to see that 
$J,J_{IJ}$ satisfy the angular momentum algebra.\\
Denote an orthonormal basis of states so constructed by 
$|j_{IJ},j;j_1,j_2,j_3>$ where it is 
understood that one first couples $j_I,j_J$ to resulting spin $j_{IJ}$
and then $j_{IJ},j_K$ to the resulting total spin $j$. Any choice
of such a recoupling scheme leads to an orthonormal basis and thus 
the transformation between these bases must be unitary. The matrix elements
of this transformation were explicitly computed by Racah. In particular 
we have
\ba \label{a}
&& <j_{12},j;j_1,j_2,j_3|j_{23},j;j_1,j_2,j_3>\nonumber\\
&=:& (-1)^{j_1+j_2+j_3+j}\sqrt{(2j_{12}+1)(2j_{23}+1)}
\left\{ \begin{array}{ccc}
j_1 & j_2 & j_{12}\\
j_3 & j   & j_{23}
\end{array} \right\}
\ea
where the $2\times 3$ matrix on the right hand side is the so-called 
$6j-$symbol for which a closed formula exists \cite{20}.\\ 
There exists a choice of phases for the basis vectors such that all the  
$6j-$symbols are real. With this choice they enjoy a large amount of 
symmetries of which we just need two : \\
1) it is invariant under an arbitrary permutation of its columns and\\
2) it is crossing-symmetric, that is
\be \label{b}
\left\{ \begin{array}{ccc}
j_1 & j_2 & j_3\\
j_4 & j_5 & j_6
\end{array} \right\} =
\left\{ \begin{array}{ccc}
j_1 & j_6 & j_5\\
j_4 & j_3 & j_2
\end{array} \right\} \;.
\ee
For the 
purposes of this paper it is sufficient to table the value of the 
$6j-$symbol for the special case that one of $j_I,j_{IJ},j$ takes the 
value $1/2$. Then we have the special values
\ba \label{c}
\left\{ \begin{array}{ccc}
a & b & c\\
\frac{1}{2} & c-\frac{1}{2} & b+\frac{1}{2}
\end{array} \right\}
&=&(-1)^{a+b+c}\sqrt{\frac{(a+c-b)(a+b-c+1)}{(2b+1)(2b+2)(2c)(2c+1)}}
\\ \label{d}
\left\{ \begin{array}{ccc}
a & b & c\\
\frac{1}{2} & c-\frac{1}{2} & b-\frac{1}{2}
\end{array} \right\}
&=&(-1)^{a+b+c}\sqrt{\frac{(a+b+c+1)(b+c-a)}{(2b)(2b+1)(2c)(2c+1)}} \;.
\ea
This is all one needs to know in order to verify the eigenvalue calculations 
of the present paper. It is clear how to generalize the recoupling theory 
for any number of spins.

\end{appendix}

\end{document}